\documentclass[aps,prx,twocolumn,superscriptaddress,noshowpacs,floatfix]{revtex4-2}

\usepackage{graphicx}
\usepackage{xcolor}
\usepackage{float}
\usepackage{bm}
\usepackage{braket}
\usepackage{amsmath}
\usepackage{lipsum}
\usepackage{soul}
\usepackage[normalem]{ulem}

\begin{document}


\title{Orbital-specific Itinerancy and Localization in a Kagome Magnet}

\author{S. V. Streltsov}
\altaffiliation {\emph{streltsov@imp.uran.ru} }
\affiliation{Institute of Metal Physics, S. Kovalevskaya Street 18, 620990 Ekaterinburg, Russia}
\affiliation{Department of theoretical physics and applied mathematics, Ural Federal University, Mira St. 19, 620002 Ekaterinburg, Russia}

\author{H. Y. Huang}
\affiliation{National Synchrotron Radiation Research Center, Hsinchu 30076, Taiwan}

\author{A. Ushakov}
\affiliation{Institute of Metal Physics, S. Kovalevskaya Street 18, 620990 Ekaterinburg, Russia}

\author{C. I. Wu}
\affiliation{Graduate Program in Science and Technology of Synchrotron Light Source, National Tsing Hua University, Hsinchu 30013, Taiwan}

\author{A. Singh}
\affiliation{National Synchrotron Radiation Research Center, Hsinchu 30076, Taiwan}

\author{J. Su}
\affiliation{National Synchrotron Radiation Research Center, Hsinchu 30076, Taiwan}

\author{J. Okamoto}
\affiliation{National Synchrotron Radiation Research Center, Hsinchu 30076, Taiwan}

\author{C. T. Chen}
\affiliation{National Synchrotron Radiation Research Center, Hsinchu 30076, Taiwan}

\author{K. Wang}
\affiliation{Keck Center for Quantum Magnetism and Department of Physics and Astronomy, Rutgers University, Piscataway, NJ 08854, USA}

\author{A. I. Poteryaev}
\affiliation{Institute of Metal Physics, S. Kovalevskaya Street 18, 620990 Ekaterinburg, Russia}
\affiliation{Department of theoretical physics and applied mathematics, Ural Federal University, Mira St. 19, 620002 Ekaterinburg, Russia}

\author{S-W. Cheong} 
\affiliation{Keck Center for Quantum Magnetism and Department of Physics and Astronomy, Rutgers University, Piscataway, NJ 08854, USA}

\author{A. Fujimori}
\affiliation{National Synchrotron Radiation Research Center, Hsinchu 30076, Taiwan}
\affiliation{Department of Physics and Center for Quantum Science and Technology, National Tsing Hua University,
Hsinchu 30013, Taiwan}
\affiliation{Department of Physics, University of Tokyo, Hongo 7-3-1, Tokyo 113-0033, Japan.}

\author{D. J. Huang}
\altaffiliation {\emph{djhuang@nsrrc.org.tw}} 
\affiliation{National Synchrotron Radiation Research Center, Hsinchu 30076, Taiwan}
\affiliation{Department of Electrophysics, National Yang Ming Chiao Tung University, Hsinchu 30010, Taiwan}
\affiliation{Department of Physics, National Tsing Hua University, Hsinchu 30013, Taiwan}

\begin{abstract}

The kagome lattice naturally hosts flat bands, Dirac fermions, and van Hove singularities, yet whether its geometry can stabilize orbital-selective phases — a hallmark of Hund's physics in multi-orbital correlated systems — has remained an open question. Here, we combine resonant inelastic X-ray scattering with density functional theory and dynamical mean-field theory to demonstrate that YMn$_6$Sn$_6$  exhibits a spontaneous orbital differentiation into coexisting itinerant and localized electrons within the same Mn $3d$ manifold. Orbitals directed along Mn-Mn bonds provide coherent quasiparticles and metallic bands, while those pointing toward ligands become strongly correlated and display non-Fermi-liquid behavior. Hund's intra-atomic exchange suppresses orbital fluctuations, stabilizing this dichotomy and providing a natural double-exchange-like mechanism for the observed ferromagnetic bilayer coupling. Our work establishes YMn$_6$Sn$_6$ as a kagome platform where orbital selectivity, flat-band topology, and Hund's metallicity converge — revealing that geometric frustration and correlation-driven orbital differentiation can cooperatively design exotic quantum phases beyond the canonical paradigms of Mott physics or band topology alone.

\end{abstract}


\maketitle

\thispagestyle{empty}



Strong electron correlations play an important role in many transition-metal compounds. Due to the significant Coulomb repulsion, electrons can become localized, resulting in Mott insulating properties. Mottness, arising from strong electron correlations, is pivotal in unraveling the enigmatic electronic and magnetic properties of correlated electron materials. For example, cuprate superconductors are proximate to a Mott insulator and have a single active electron band at the Fermi level. 

Recent advances in the study of kagome crystals have revealed intriguing connections between many-body electron correlations and topological quantum effects~\cite{kang2020b,nag2022,li2025}. A kagome lattice is a two-dimensional crystal structure composed of corner-sharing triangles, and features three key properties: Dirac fermions, flat bands and van Hove singularities in its electronic structure. The combination of strong electronic correlations and non-trivial band topology can lead to instabilities, such as charge density waves \cite{Kiesel2013}, superconductivity \cite{zhao2021,yin2022}, or magnetic instability \cite{ma2022,sales2022}. The kagome magnet YMn$_6$Sn$_6$ has double Mn layers in which Mn spins are ferromagnetically coupled and rotate from one bilayer to the next along the $c$-axis, forming a helical spin structure close to room temperature~\cite{Yoshimura1983,Ghimire2020}.  It has a complex phase diagram with magnetic field-induced Lifshitz transition~\cite{Siegfried2022}, and demonstrates a large anomalous transverse thermoelectric effect~\cite{Roychowdhury2022} as well as helicity-controlled non-reciprocal transport~\cite{Yamauchi2025}. YMn$_6$Sn$_6$ was argued to be a Hund's metal with the kagome lattice~\cite{li2021}; it is expected to exhibit exotic electronic properties arising from a combined effect of topologically nontrivial multi-orbitals with strong electron correlations. 

 The interplay between orbital degeneracy and electron correlations can result in complex electronic and magnetic phases. For example, multi-orbital systems with both Hubbard $U$ and Hund's intra-atomic exchange $J_{\rm H}$ can undergo an orbital-selective Mott transition~\cite{Anisimov2002a}; the presence of Hund's exchange influences which orbitals remain conducting and which ones become localized~\cite{Medici2009}.  In addition, the intra-atomic exchange interaction of spins can effectively alter the repulsive Coulomb energy. For a half-filled shell, the Hund's coupling favors the electronic configurations with parallel spins in different orbitals. Hence, the effective Coulomb repulsion is increased by the Hund's coupling, while it is reduced for other fillings~\cite{khomskii2014transition}. The Hund's coupling also suppresses orbital fluctuations and stabilizes metallic states close to the half filling, leading to the emergence of a distinct type of metal known as Hund's metal \cite{Yin2011,Georges2013}. Hund's metals are not in close proximity to a Mott insulating state but have signatures of strong correlations; they are strongly correlated but at the same time have a character specific to itinerant systems. It is, therefore, an intriguing problem how Hund's metal behaviors manifest themselves in topologically nontrivial kagome magnets.

In this article, we present resonant inelastic X-ray scattering (RIXS) measurements and first-principles calculations using density functional theory (DFT) and dynamical mean-field theory (DMFT) for the kagome magnet YMn$_6$Sn$_6$. We observed unusual RIXS excitations consisting of Raman- and fluorescence-like excitations in incident-photon-energy dependent measurements. The DFT+DMFT calculations revealed strong electron correlations in this multi-orbital and topologically nontrivial magnet, indicating orbital selectivity stabilized by Hund's exchange.

\vspace{5mm}
\noindent{\bf RESULTS}

\vspace{3mm}
\noindent\textbf{RIXS excitations}
\vspace{2mm}

\begin{figure}[t!]
\centering
\includegraphics[width=1\columnwidth]{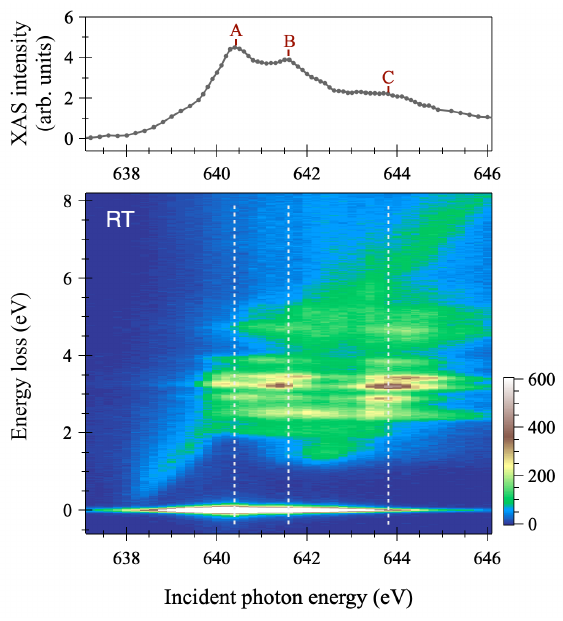}
\caption{Mn $L_3$-edge XAS and RIXS of YMn$_6$Sn$_6$. (a) Mn $L_3$-edge XAS data recorded at a temperature of 300~K. (b) RIXS intensity map plotted in the plane of energy loss vs. incident X-ray energy. The energy of incident $\pi$-polarized X-ray was tuned across the $L_3$-edge. The incident and scattering angles were 20 and 90 degrees, respectively.} 
\label{RIXS1}
\end{figure}

RIXS is a powerful spectroscopy technique used to probe low-energy excitations with momentum resolution.  When X-rays shine onto a material, some of the incident X-rays are scattered by the sample, and lose energy through various processes, such as the excitation of electrons to higher energy levels or the creation of collective excitations like phonons or magnons. Energy loss and momentum transfer are measured and analyzed to gain insights into the dynamic processes occurring within the sample. This versatile method proves invaluable for detecting a wide range of excitation phenomena, including local and collective excitations such as $d-d$ excitations, charge transfers, phonons, and spin excitations. One notable feature of these RIXS excitations is that the energy loss is independent of the incident photon energy, known as Raman-like RIXS. Additionally, RIXS probes intra- or inter-band transitions, yielding fluorescence-like RIXS signatures, where the energy loss increases with the increase in incident photon energy. 

We performed RIXS measurements on YMn$_6$Sn$_6$ using various incident photon energies across the Mn $L_3$-edge. Figure~\ref{RIXS1} displays the RIXS intensity map in the plane defined by energy loss and incident photon energy. This intensity map reveals two distinct RIXS signatures: fluorescence-like and Raman-like RIXS features. For energy loss between 2 and 5.5 eV, the RIXS spectral weight is dominated by Raman-like excitations originating from Mn $d-d$ excitations, whose energies are independent of incident photon energy. Similar to the Raman-to-fluorescence crossover observed in other transition-metal compounds~\cite{Ghiringhelli2007,Zhou2011,Bisogni2016,Hariki2018,Gilmore2021}, we also observed fluorescence-like signatures on energy above the $d-d$ excitations, arising from the emission from occupied valence states. In addition, the measured RIXS also shows weak fluorescence-like features below the $2p\rightarrow3d$ resonance, reflecting the transitions of $3d$ bands crossing the Fermi level. The low-energy fluorescence-like RIXS feature has been attributed to distinct mechanisms across different compounds~\cite{Zhou2011,Bisogni2016,Hariki2018}. It reflects either unbound particle–hole excitations in the RIXS final state or the presence of itinerant carriers. In short, our RIXS results reveal that YMn$_6$Sn$_6$ exhibits localized and delocalized characteristics of excitation.

\vspace{3mm}
\noindent\textbf{Multiplet calculations}
\begin{figure}[t!]
\centering
\includegraphics[width=1\columnwidth]{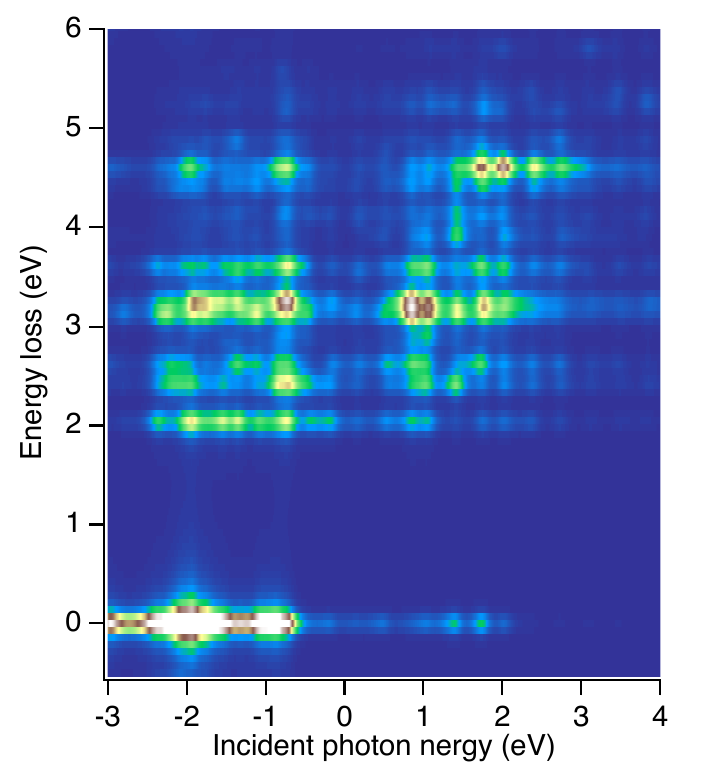}
\caption{{\bf Calculated RIXS spectra obtained using ligand-field multiplet calculations.}
The calculations were carried out with $Quanty$ within an octahedral ($O_h$) crystal-field symmetry, characterized by a crystal-field splitting parameter of 10$Dq = 0.45$ eV. The intra-atomic exchange interaction (Hund’s coupling) was included through the Slater integrals $F^2$ and $F^4$, which were reduced to 78\% of their Hartree–Fock values to account for configuration-interaction effects. The charge-transfer energy was set to $\Delta = 2.5$ eV. The hybridization strength between Mn $3d$ and ligand $2p$ orbitals was described by the Slater–Koster parameters ($pd\sigma$ and $pd\pi$), with hybridization energies $V_{e_g} = 2.06$ eV and $V_{t_{2g}} = 1.21$ eV.
}
\label{rixs_cal}
\end{figure}

To understand the electronic structure of the Mn $3d$ associated with the observed dual RIXS signatures, we used ligand-field multiplet calculations to analyze the measured RIXS spectra. In these calculations, the ground state is expressed as an expansion of the crystal field configurations $3d^n$ and of the charge-transfer configurations $3d^{n+1}\underline{L}$ with $n=5$ for Mn$^{2+}$ and $\underline{L}$ being a ligand hole. With the crystal field $10Dq$ and charge-transfer energy $\Delta$ deduced from the first-principles DFT calculations as the starting point~\cite{mahadevan1996}, we calculated RIXS spectra at various incident photon energies across the Mn $L_3$-edge. Since the measured RIXS spectra of the $d$–$d$ excitations resemble those of MnO \cite{Ghiringhelli2006}, we adopted a crystal field of $O_h$ symmetry to simplify the ligand-field multiplet calculations. Figure~\ref{rixs_cal} shows the theoretical RIXS intensity map of Mn$^{2+}$ and reproduces the $d-d$ excitation of energies independent of incident photon energy, demonstrating the localized character of Mn $3d$ states. However, as expected, multiplet calculations are unable to explain fluorescence-like excitations.

\vspace{3mm}
\noindent\textbf{DFT calculations}

We turn to analysis of {\it ab initio} DFT and DFT+DMFT calculation results, which can be used to describe both local many-body electron correlations and formation of the band structure.

\begin{figure}[t!]
\centering
\includegraphics[width=1\columnwidth]{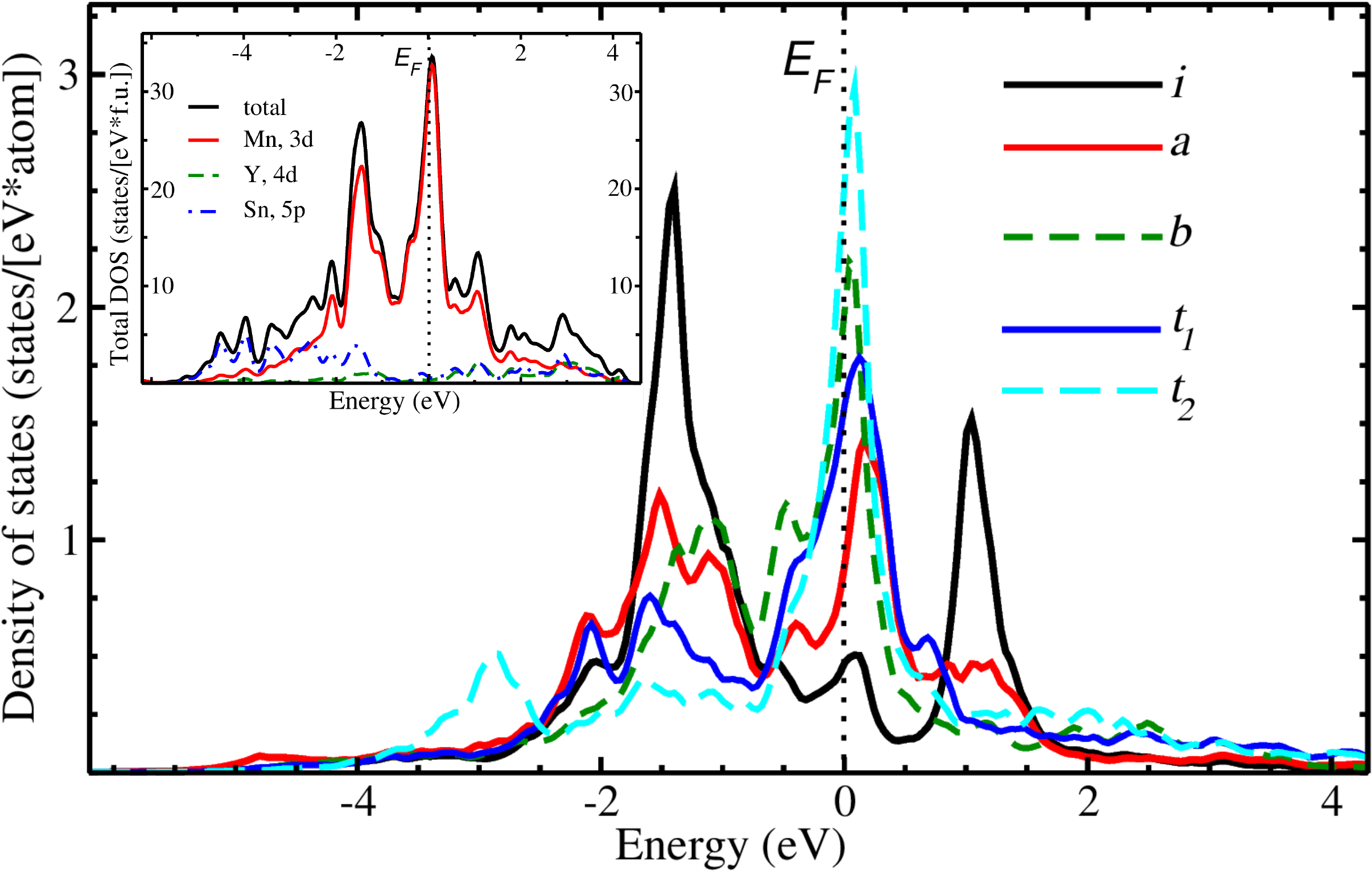}
\caption{Partial densities of states (DOS) for Mn $3d$ orbitals in non-magnetic DFT calculations (see figure for color coding). Inset shows Mn 3$d$, Y 4$d$ and Sn 5$p$, by red, green and blue colors, respectively. The Fermi level is at zero. 
}
\label{DOS_DFT}
\end{figure}

The total and partial densities of states (DOS) are shown in Figure~\ref{DOS_DFT}. The flat band specific for the kagome lattice and observed in the ARPES measurement~\cite{li2021} is seen as a shoulder in total DOS or small peak in partial DOS at $-0.4$~eV. Our Wannier function projection of DFT Hamiltonian~\cite{korotin2008construction} shows that the crystal-field splitting due to Sn and Y ions is moderate and the difference between the lowest and highest eigenvalues is only 810~meV. Indeed, the average Mn-Sn distance of $2.80$~\AA~is much larger than that in typical oxides, where the Mn-O distance is about 2.0~{\AA} in manganites. Moreover, the Mn-Mn distance of $2.76$~\AA~is even shorter than Mn-Sn bond, and it is mush smaller than Mn-Mn distances in manganites ($\sim 4.0$~{\AA}). Thus, the effect of the electric field (and hybridization) of neighboring Mn ions is even more important. These positively charged Mn ions form a rectangular surrounding and lower the energy of the electronic orbitals directed along Mn-Mn bonds. 

\begin{figure}[t!]
\centering
\includegraphics[width=1\columnwidth]{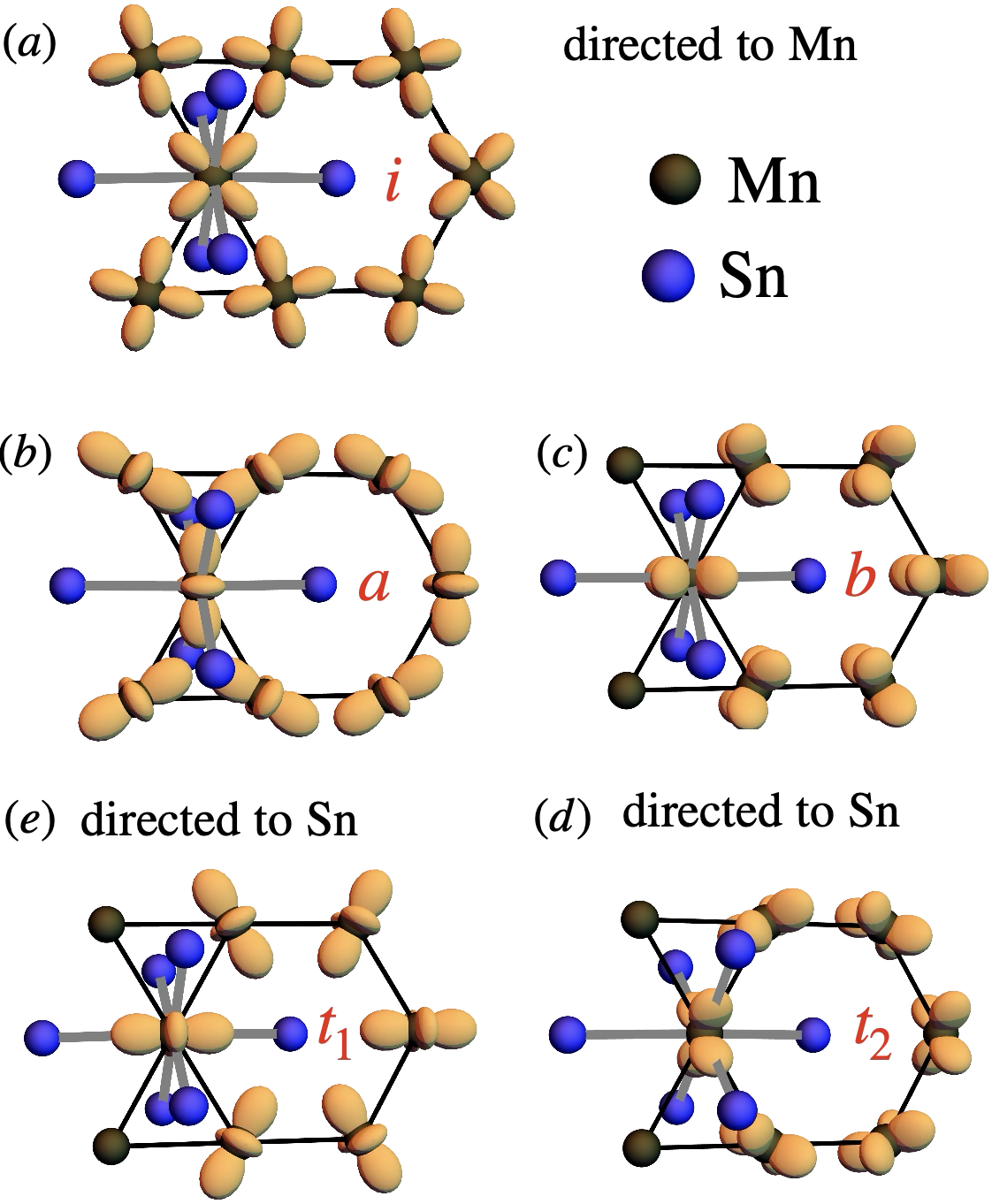}
\caption{(a)-(e) Mn $3d$ orbitals obtained by the diagonalization of the local part of the corresponding DFT Hamiltonian. Mn atoms are shown by dark gray and Sn by violet balls. One can see that $i$ orbital is directed along Mn-Mn bonds, in contrast $t_{1,2}$ orbitals look mostly to Sn.}
\label{orbitals}
\end{figure}

In low-symmetry positions of Mn ions, $C_{2v}$ point group, degeneracy of $d$ shell is lifted completely. Diagonalization of the $d$-$d$ block of the Wannier function projected DFT Hamiltonian used in subsequent DFT+DMFT calculations gives further insight into the details of the electronic structure. The corresponding orbitals are plotted in Figures~\ref{orbitals}(a)-(e), and their contributions to DOS are shown in Figure~\ref{DOS_DFT}.  One orbital, which we label as $i$, has its lobes directed as much as possible along the Mn-Mn bonds. The $a$ and $b$ orbitals are similar to $i$ in that they also point away from the Sn ions. The final two orbitals,  $t_1$ and $t_2$, are directed toward the ligand atoms. Although hybridization with Y $3d$ and Sn $5p$ states, as well as off-diagonal elements between these states at different $k$-points, makes the situation complex, one can clearly see that the electrons in the $i$ band are more itinerant and that its bandwidth is significantly larger (see Figure~\ref{DOS_DFT}). In the next section, we will demonstrate that this ``orbital differentiation'' is further enhanced by correlation effects, particularly by the Hund's intra-atomic exchange interaction.

\vspace{3mm}
\noindent\textbf{Orbital-selective electron correlations}

We applied the DMFT approach~\cite{Anisimov97,Lichtenstein1998} to take into account strong correlations lying beyond DFT. 
The DFT+DMFT calculation scheme has been previously used in Ref.~\cite{li2021} for the interpretation of the angle-resolved photoemission data of the present material, and the DMFT results also evidence a strong fluctuation of the magnetic moment, which was ascribed to Hund's physics.  In the following, we shall first present the DMFT results for the paramagnetic state, 
and then for the ferromagnetic phase, which better represent a helical magnetic order at low temperature.

\begin{table}[b] 
\centering
\caption{Local (instant) magnetic moment measured as $m_z = \sqrt{\langle M^2_z \rangle}$ and probability of different electronic configurations on Mn ions for different sets of Hubbard $U$ and Hund's $J_{\rm H}$ (given in eV) obtained in the paramagnetic DFT+DMFT calculations at $T=386$~K.}
\begin{tabular}{l  c c c c c }
\hline
\hline
\multicolumn {1}{c}{}& \multicolumn {3}{c}{$U=3.75$} & &$U=2.75$ \\
  $J_{\rm H}$      & $0.71$  & $0.81$ & $0.91$ && $0.81$\\
\hline
 $m_z$ & 3.13$\mu_B$& 3.49$\mu_B$ &  3.85$\mu_B$ && 3.34$\mu_B$\\
$d^4$ &    4\% & 4\% & 5\% && 5\%\\
 $d^5$ &  26\% & 31\% & 39\% && 30\%\\
 $d^6$ &  48\% & 47\% & 44\% && 44\%\\
 $d^7$ &  18\% & 15\%  & 10\% && 17\%\\
\hline
\hline
\end{tabular}
\label{tab:conf-table}
\end{table}

Paramagnetic DMFT calculations show that the local magnetic moment (estimated as $\sqrt{\langle M^2_z \rangle}$) is somewhat smaller than 5$\mu_B$ expected for the high-spin $d^5$ ($S=5/2$) state of Mn, but more importantly, it strongly depends on the Hund's coupling parameter $J_{\rm H}$ as one can see from Table~\ref{tab:conf-table}. The Hund's coupling not only influences longitudinal spin fluctuations but is also detrimental to charge fluctuations. As shown in Table~\ref{tab:conf-table}, the Mn ion is mainly in the $d^6$ electronic configuration for $J_{\rm H}=0.71$~eV, but the weight of the $d^5$ configuration rapidly grows with $J_{\rm H}$. Interestingly, Hubbard $U$ has little effect on the probabilities of different charge states and only moderately changes the local magnetic moment. It is important to note the interaction parameters estimated by the linear response theory~\cite{cococcioni2005} are $U=3.75$~eV and $J_{\rm H}=0.81$~eV.

A fundamental difference between DMFT and the ligand-field multiplet calculations described in one of the previous sections is that DMFT considers localized and band electrons on an equal footing. Remarkably, the difference between the Mn $3d$ orbitals, which is clearly seen in DFT calculations, not only persists when correlation effects are included in DMFT but it is even enhanced by Hund's coupling. Indeed, already DMFT calculations for the paramagnetic phase reveal a pronounced orbital selectivity in the imaginary part of the self-energy, $\Sigma (i\omega_n)$. Within the Fermi-liquid framework, Im$\Sigma(i\omega)$ should decrease linearly to zero at small Matsubara frequencies, whereas its divergence indicates an insulating behavior.  As clearly shown in Figure S1~\cite{supp}, $\rm {Im} \Sigma (i\omega_n)$ diverges at small $i \omega_n$ for the localized $t$ orbitals (pointing towards ligands), whereas the $i$ orbitals (directed along Mn-Mn bonds) have a Fermi-liquid character. The $a$ and $b$ orbitals can exhibit either character—Fermi-liquid or non-Fermi-liquid—depending on the value of $J_H$.

\begin{figure}[t!]
\centering
\includegraphics[width=1\columnwidth]{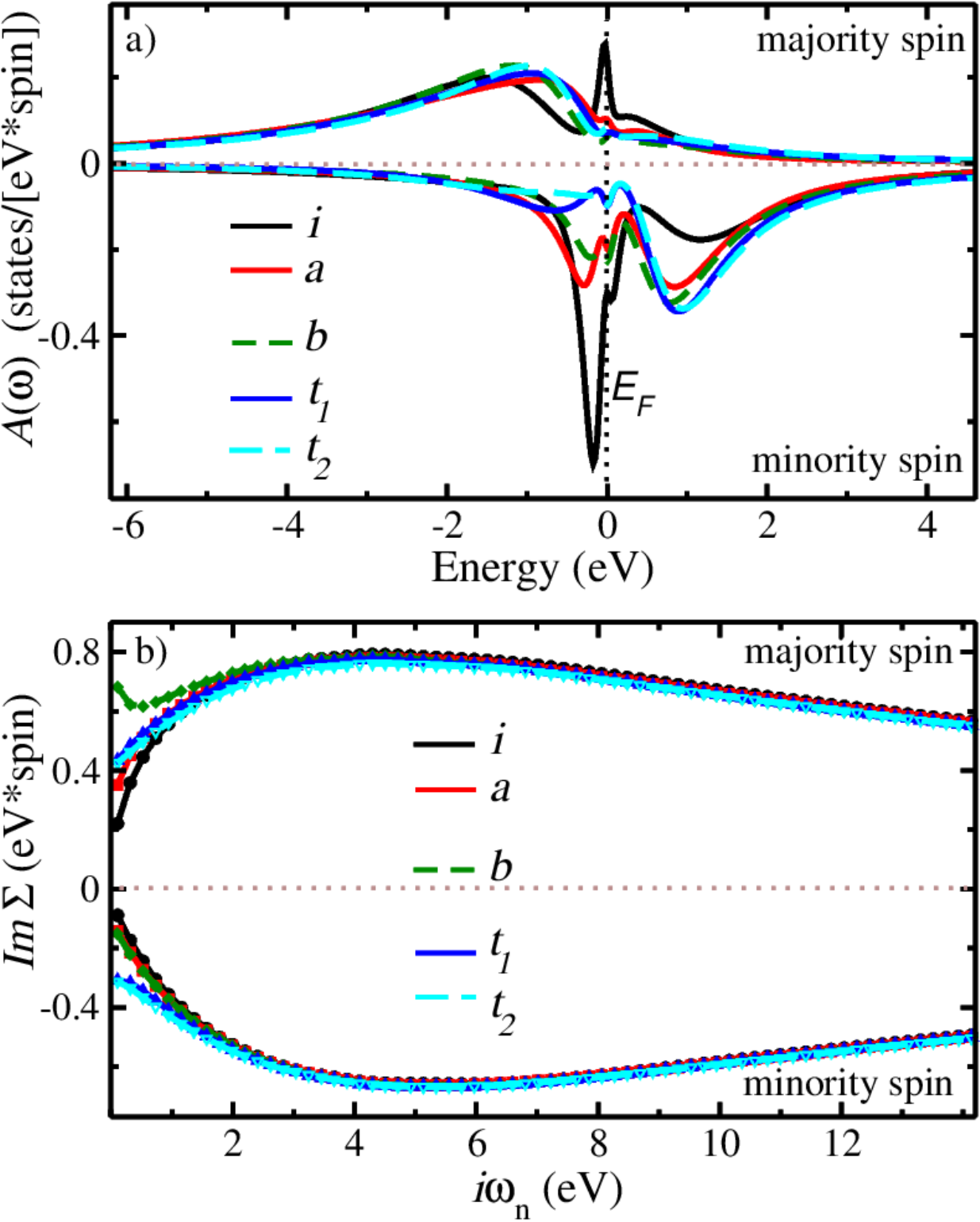}
\caption{Results of the ferromagnetic DFT+DMFT calculation at $U=3.75$~eV, $J_{\rm H}=0.81$~eV, and $T=386$~K. a) --  spin and orbital resolved spectral functions for Mn 3$d$ states; b) -- corresponding imaginary parts of self-energy. We note, that since RIXS is described by two-particle processes, these spectral functions can not be directly compared with the experimental results.
}
\label{fig:spectral-functions}
\end{figure}

Very similar effects are observed in the magnetically ordered phase. Because the ferromagnetic coupling between Mn atoms within each layer is considered to be much stronger than that between layers and the ferromagnetic state approximately represents the helical-spin state, we performed ferromagnetic DFT+DMFT calculations.
They show that the system prefers to have spontaneous magnetization for $T \lesssim 500$~K. The factor of 1.5 overestimation of the N\'eel temperature (experimental $T_{N} = 345$~K~\cite{Ghimire2020}) is rather typical for such mean-field methods as DMFT~\cite{Anisimov2012}.


Analysis of the spin- and orbital-resolved spectral functions $A(\omega)$ and the frequency dependence of the imaginary part of self-energy Im$\Sigma(i\omega)$ presented in Figures~\ref{fig:spectral-functions}a and \ref{fig:spectral-functions}b, respectively, gives insights into the effects of correlations in YMn$_6$Sn$_6$. In the Fermi-liquid regime, $A(\omega)$ exhibits a sharp, coherent peak near the Fermi level and a broad, incoherent spectral features at higher energies. When electronic correlations are strong, the quasiparticle peak is suppressed, and spectral weight is transferred from the coherent peak to the incoherent part~\cite{Fujimori1992,DJH2003,Perfetti2003,Takizawa2009}. The $A(\omega)$ of the $i$ orbital is dominated by a coherent peak. This is consistent with the fluorescence-like interband excitations observed experimentally.  The incoherent part of $A(\omega)$, i.e., the broad peaks at $\omega\sim -1$ and 1 eV dominating the spectral function, represents the nearly localized part of the Mn $3d$ electrons, and is likely responsible for the $d$-$d$ excitations in our RIXS spectra. 

Similar to the paramagnetic phase, orbital-selective behaviour is also observed in the magnetically ordered phase. Figure~\ref{fig:spectral-functions}b shows that in the minority-spin channel: the $i$ and $a,b$ orbitals exhibit metallic-like behavior with a small imaginary part of the self-energy at the first Matsubara point, while Im$\Sigma(i\omega)$ for the $t$ orbitals appears insulating, resulting in a strong suppression of the quasi-particle peaks at the Fermi level and transferring the spectral weight to higher energies. Smaller values of the imaginary parts for the $i$ state at low Matsubara frequencies lead to the large quasi-particle peak at the Fermi level shown in Figure~\ref{fig:spectral-functions}a. Spin majority states of Mn are nearly completely filled and, therefore, less interesting, but orbital-selectivity is clearly observed in them too.


\vspace{3mm}
\noindent{\bf DISCUSSION}

The present analysis naturally explains the anomalous RIXS spectrum discussed above. We see that both theoretical DFT+DMFT calculations and RIXS measurements demonstrate that YMn$_6$Sn$_6$ is in the orbital selective regime with metallic-like $i$ electrons, and more localized $a,b$, and especially $t$ electrons. This immediately suggests that the experimentally observed dominant ferromagnetic interaction in the Mn bi-layer can be explained by the double exchange mechanism~\cite{khomskii2014transition}. This agrees with conclusions of Ref.~\cite{Ghimire2020,biswas2025}, where the Ruderman–Kittel–Kasuya–Yosida (RKKY) mechanism of exchange interaction was suggested (double exchange is a special case of RKKY), but care should be taken, since there can also be contributions coming from the super-exchange interaction via Sn-$p$ orbitals.

By the combination of RIXS spectroscopy and theoretical DFT and DFT+DMFT calculations on  YMn$_6$Sn$_6$, we identified the new aspect of the kagome lattice related to a substantial orbital selectivity, supported by Hund's intra-atomic exchange interaction. The $i$ orbitals directed along the Mn-Mn bonds on the kagome lattice strongly overlap, which results in a large bandwidth and suggest an itinerant character of these electrons. The intra-atomic Hund's coupling suppresses orbital fluctuations and, therefore, stabilizes the orbital-selective regime when we go from a metal phase as it was shown in ~\cite{tocchio2016assessing,Georges2013}. This is exactly what we see in the DMFT calculations for YMn$_6$Sn$_6$: at critical $J_{\rm H} \sim 0.8$ eV there appear two types of carriers: itinerant $i$ electrons which form quasi-particle peaks, localized $t$ electrons and $a,b$ states with non-Fermi-liquid behavior. This observation is analogous to the double exchange mechanism in manganites, in which the Hund's coupling is essential and the hopping of cubic $e_g$ electrons is most efficient when the spins of the Mn ions are parallel. While the strong Hund's coupling in the double exchange mechanism aligns the $e_g$ electron spin with the localized $t_{2g}$ spins in manganites, the orbital-selectivity can explain the ferromagnetic coupling in the Mn bi-layers of YMn$_6$Sn$_6$. Additionally, the observed dual characteristic in the electronic excitation of YMn$_6$Sn$_6$ agrees with the Mn $2p$ photoemission results of La$_{1-x}$Sr$_x$MnO$_3$, which display both well-screened and multiplet peaks~\cite{Horiba2004,Uozumi1997}. 

The orbital-selective behavior observed in this study is likely common to many other kagome magnets. These materials share not only the same lattice structure but also the key features responsible for orbital selectivity: short metal-metal bonds and being composed of $3d$ transition metals with similar intra-atomic Hund's exchange.

\vspace{5mm}
\noindent{\bf METHODS}

\noindent{\bf Sample synthesis}

\noindent Single crystals of YMn$_6$Sn$_6$ were grown by using high-temperature flux method with excess Sn as flux. Y piece (Alfa Aesar, purity 99.9\%), Mn granules (Alfa Aesar, purity 99.9\%), and Sn shots (Alfa Aesar, purity 99.99\%) with a molar ratio of Y:Mn:Sn~=~1:6:30 were put into a 2-ml Canfield alumina crucible set and sealed in a quartz ampoule under partial argon atmosphere. The sealed quartz ampoule was heated up to 1273~K and held for 24~h. Then it was cooled down slowly to 873~K at a rate of 3~K/h, at which the ampoule was taken out from the furnace and decanted with a centrifuge to separate YMn$_6$Sn$_6$ crystals from excess Sn flux leaving behind well-faceted hexagonal crystals. The crystal structure of the compound was verified by X-ray powder diffraction at room temperature using a Bruker D2 diffractometer. A few crystals from each growth batch were ground into powder, and X-ray diffraction patterns were collected on those powder samples. Magnetization and electrical transport measurements were carried out by using Quantum Design PPMS-9 T.

\vspace{3mm}
\noindent{\bf RIXS measurements}

\noindent We conducted Mn $L$-edge RIXS measurements using the AGM-AGS spectrometer at beamline 41A of the Taiwan Photon Source, National Synchrotron Radiation Research Center, Taiwan \cite{SinghJSR2021}. The AGM-AGS beamline is designed based on the energy-compensation principle of grating dispersion. The scattering angle was fixed at 150$^\circ$, and the RIXS scattering plane was defined by the [100] and [001] directions in reciprocal space. Prior to the RIXS measurements, the incident photon energy was calibrated using X-ray absorption spectra recorded in the fluorescence-yield mode with a photodiode. The total energy resolution of the RIXS setup, including contributions from both the monochromator and the spectrometer, was 30~meV (full width at half maximum) at an incident photon energy of 630~eV. The RIXS spectra were collected using a CMOS image detector without analyzing the polarization of the scattered X-rays. The measurements were performed at room temperature.

\vspace{5mm}
\noindent\textbf{Calculation details}

\noindent The crystal structure of YMn$_6$Sn$_6$ at ambient pressure was taken from Ref.~[\onlinecite{Romaka-11}]. DFT calculations were performed using the pseudopotential method as realized in the Quantum Espresso code~\cite{Gianozzi-09}. The exchange-correlation potential was taken in the form proposed by Perdew {\it et al.}~\cite{Perdew-96}. The $k$-grid consisted of 700 points in the whole Brillouin zone, and the wavefunction cutoff was chosen to be 40 Ry. The values of the onsite Coulomb repulsion parameter, $U$, and Hund's coupling constant, $J_{\rm H}$, were calculated to be $U = 3.75$~eV and $J_{\rm H} = 0.81$~eV for Mn-$3d$ electrons using the linear response approach~\cite{cococcioni2005}. The crystal-field splitting on Mn sites was estimated using Wannier function projection\cite{korotin2008construction} on Mn $3d$ states only. 

A non-interacting DFT Hamiltonian including the Mn $3d$, Sn $5p$, and Y $4d$ states for DMFT calculations was generated using the Wannier projection procedure~\cite{korotin2008construction}. In order to exclude the $d-d$ interaction taken into account on the DFT level of approximation, we used a double-counting correction calculated as $\hat{H}_{dc} = \widetilde{U}(n_{dmft}-1/2)\hat{I}$. Here, $n_{dmft}$ is the self-consistent total number of correlated $d$ electrons obtained within the DFT+DMFT approach, $\widetilde{U}$ is the average Coulomb parameter for the $d$-shell, $\hat{I}$ is the unit operator.
The impurity solver used in DMFT calculations was based on the segment version of the hybridization expansion Continuous Time Quantum Monte-Carlo method (CT-QMC)~\cite{Werner-06}.

\vspace{5mm}
\noindent\textbf{Data Availability}

\noindent All data generated or analyzed during this study are available from the corresponding authors upon reasonable request.

\vspace{4mm}
\noindent\textbf{Acknowledgments}

\noindent S.V.S. thanks T. Saha-Dasgupta, M. Gr\"uninger, and I. Mazin for fruitful discussions. This work was supported in part by the Ministry of Science and Technology of Taiwan under Grant Nos. 109-2112-M-213-010-MY3, 109-2923-M-213-001, and 113-2112-M-007-033, and by the Japan Society for the Promotion of Science under Grant No. JP22K03535.  The work at Rutgers University was supported by the DOE under Grant No. DOE: DE-FG02-07ER46382. S.V.S., A.V.U., and A.I.P. acknowledge support of the Ministry of Science and Higher Education of the Russian Federation through the IMP UB RAS. A.F. acknowledges the support from the Yushan Fellow Program under the Ministry of Education of Taiwan.


\vspace{4mm}
\noindent\textbf{Author Contributions}

\noindent D.J.H. and S.V.S. conceived and coordinated the project. H.Y.H., C.I.W., A.S., J.S., J.O., and C.T.C. conducted the RIXS experiments. K.W. and S.W.C. synthesized and characterized the samples. A.V.U. and S.V.S. performed DFT calculations, while DFT+DMFT calculations were carried by A.V.U., S.V.S., and A.I.P. D.J.H., S.V.S., and A.F. wrote the manuscript with inputs of other authors.

\vspace{4mm}
\noindent\textbf{Competing interests}

\noindent The authors declare that there are no competing interests.

\bibliography{ref}

@article{SinghJSR2021,
  title={{Development of the soft X-ray AGM-AGS RIXS beamline at the Taiwan photon source}},
  author={Singh, A and Huang, H. Y. and Chu, Y. Y. and Hua, C. Y. and Lin, S. W. and Fung, H. S. and Shiu, H. W. and Chang, J and Li, J. H. and Okamoto, J and Chiu, C. C. and Chang, C. H. and Wu, W. B. and Perng, S. Y. and Chung, S. C. and Kao, K. Y. and Yeh, S. C. and Chao, H. Y. and Chen, J. H. and Huang, D. J. and Chen, C. T.},
  journal={J.Synchrotron Radiat.},
  volume={28},
  number={3},
  pages={977-986},
  year={2021},
  publisher={International Union of Crystallography}
}

@article{Georges2013,
  title={Strong correlations from Hund’s coupling},
  author={Georges, Antoine and Medici, Luca de' and Mravlje, Jernej},
  journal={Annu. Rev. Condens. Matter Phys.},
  volume={4},
  number={1},
  pages={137--178},
  year={2013},
  publisher={Annual Reviews}
}

@article{Anisimov2002a,
   author = {V.I. Anisimov and I.A. Nekrasov and D.E. Kondakov and T.M. Rice and M. Sigrist},
   doi = {10.1140/epjb/e20020021},
   issn = {1434-6028},
   issue = {2},
   journal = {The European Physical Journal B},
   month = {2},
   pages = {191},
   title = {{Orbital-selective Mott-insulator transition in Ca$_{2-x}$Sr$_x$RuO$_4$}},
   volume = {25},
   url = {http://www.springerlink.com/index/10.1140/epjb/e20020021},
   year = {2002},
}

@article{Medici2009,
  title={Orbital-selective Mott transition out of band degeneracy lifting},
  author={de’Medici, Luca and Hassan, Syed R and Capone, Massimo and Dai, Xi},
  journal={Physical Review Letters},
  volume={102},
  number={12},
  pages={126401},
  year={2009},
  publisher={APS}
}

@article{Anisimov97,
   author = {V. I. Anisimov and A. I. Poteryaev and M. A. Korotin and A. O. Anokhin and G. Kotliar},
   journal = {Journal of Physics: Condensed Matter},
   pages = {7359},
   title = {First-principles calculations of the electronic structure and spectra of strongly correlated systems: dynamical mean-field theory},
   volume = {9},
   url = {http://iopscience.iop.org/0953-8984/9/35/010},
   year = {1997},
}

@book{khomskii2014transition,
   author = {D. I. Khomskii},
   isbn = {9781107020177},
   publisher = {Cambridge University Press},
   title = {Transition Metal Compounds},
   year = {2014},
}

@article{korotin2008construction,
  title={Construction and solution of a Wannier-functions based Hamiltonian in the pseudopotential plane-wave framework for strongly correlated materials},
  author={Korotin, Dm and Kozhevnikov, AV and Skornyakov, SL and Leonov, I and Binggeli, N and Anisimov, VI and Trimarchi, G},
  journal={The European Physical Journal B},
  volume={65},
  pages={91--98},
  year={2008},
  publisher={Springer}
}

@article{Gianozzi-09,
  title={},
  author={P. Gianozzi and S. Baroni and N. Bonini and M. Calandra and R. Car and C. Cavazzoni and D. Ceresoli and G. L. Chiarotti and M. Cococcioli and S. Fabris and G. Fratesi and R. Gebauer and U. Gerstmann and C. Gougoussis and A. Kakolj and M. Lazzeri and L. Martin-Samos},
  journal= {J. Phys. C: Condens. Matter},
  volume={21},
  pages={395502},
  year={2009}
}

@article{Perdew-96,
  title={},
  author={J. P. Perdew and K. Burke and M. Ernzerhof},
  journal={Physical Review Letters},
  volume={77},
  pages={3865},
  year={1996}
}

@article{Werner-06,
  title={},
  author={ P. Werner and A. Comanac and L. de’Medici and M. Troyer and
A. J. Millis},
  journal={Physical Review Letters},
  volume={97},
  pages={076405},
  year={2006}
}

@article{Romaka-11,
  title={Peculiarity of component interaction in {Y, Dy}–{Mn}–{Sn} ternary systems},
  author={V. V. Romaka and M. Konyk and L. Romaka and V. Pavlyuk and H. Ehrenberg and A. Tkachuk},
  journal={Journal of Alloys and Compounds},
  volume={509},
  pages={7559--7564},
  year={2011}
}

@article{tocchio2016assessing,
  title={Assessing the orbital selective Mott transition with variational wave functions},
  author={Tocchio, Luca F and Arrigoni, Federico and Sorella, Sandro and Becca, Federico},
  journal={Journal of Physics: Condensed Matter},
  volume={28},
  number={10},
  pages={105602},
  year={2016},
  publisher={IOP Publishing}
}

@article{li2021,
  title={{Dirac cone, flat band and saddle point in kagome magnet YMn$_6$Sn$_6$}},
  author={Li, Man and Wang, Qi and Wang, Guangwei and Yuan, Zhihong and Song, Wenhua and Lou, Rui and Liu, Zhengtai and Huang, Yaobo and Liu, Zhonghao and Lei, Hechang and Yin, Zhiping and Wang, Shancai},
  journal={Nature Communications},
  volume={12},
  number={1},
  pages={3129},
  year={2021},
  publisher={Nature Publishing Group UK London}
}

@article{cococcioni2005,
  title={{Linear response approach to the calculation of the effective interaction parameters in the {LDA}+{U} method}},
  author={Cococcioni, Matteo and De Gironcoli, Stefano},
  journal={Physical Review B},
  volume={71},
  number={3},
  pages={035105},
  year={2005},
  publisher={APS}
}

@article{Yin2011,
  title={Kinetic frustration and the nature of the magnetic and paramagnetic states in iron pnictides and iron chalcogenides},
  author={Yin, Z. P. and Haule, Kristjan and Kotliar, Gabriel},
  journal={Nature Materials},
  volume={10},
  number={12},
  pages={932--935},
  year={2011},
  publisher={Nature Publishing Group UK London}
}

@article{Kiesel2013,
  title = {{Unconventional Fermi Surface Instabilities in the Kagome Hubbard Model}},
  author = {Kiesel, Maximilian L. and Platt, Christian and Thomale, Ronny},
  journal = {Phys. Rev. Lett.},
  volume = {110},
  issue = {12},
  pages = {126405},
  numpages = {5},
  year = {2013},
  month = {Mar},
  publisher = {American Physical Society},
  doi = {10.1103/PhysRevLett.110.126405}
}

@article{zhao2021,
  title={{Cascade of correlated electron states in the kagome superconductor {Cs}V$_3${Sb}$_5$}},
  author={Zhao, He and Li, Hong and Ortiz, Brenden R and Teicher, Samuel ML and Park, Takamori and Ye, Mengxing and Wang, Ziqiang and Balents, Leon and Wilson, Stephen D and Zeljkovic, Ilija},
  journal={Nature},
  volume={599},
  number={7884},
  pages={216--221},
  year={2021},
  publisher={Nature Publishing Group UK London}
}

@article{mahadevan1996,
  title={{Estimates of electronic interaction parameters for LaMO$_3$ compounds (M=Ti–Ni) from ab initio approaches}},
  author={Priya Mahadevan and N. Shanthi and D. D. Sarma},
  journal={Phys. Rev. B},
  volume={54},
  number={16},
  pages={11199--11206},
  year={1996},
  doi={10.1103/PhysRevB.54.11199}
}

@article{Ghimire2020,
  title = {Competing Magnetic Phases and Fluctuation-Driven Scalar Spin Chirality in the Kagome Metal {YMn}$_6${Sn}$_6$},
  author = {Ghimire, Nirmal J. and Dally, Rebecca L. and Poudel, L. and Jones, D. C. and Michel, D. and Magar, N. Thapa and Bleuel, M. and McGuire, Michael A. and Jiang, J. S. and Mitchell, J. F. and Lynn, Jeffrey W. and Mazin, I. I.},
  year = {2020},
  month = dec,
  journal = {Science Advances},
  volume = {6},
  number = {51},
  pages = {eabe2680},
  issn = {2375-2548},
  doi = {10.1126/sciadv.abe2680},
  urldate = {2024-09-14},
  copyright = {https://creativecommons.org/licenses/by-nc/4.0/},
  langid = {english}
}

@article{Yoshimura1983,
  title = {{NMR} Study of Magnetic State of {Mn} in {Y-Mn} Intermetallic Compounds: {YMn}$_2$, {Y}$_6${Mn}$_{23}$ and {YMn}$_{12}$},
  shorttitle = {NMR Study of Magnetic State of Mn in Y-Mn Intermetallic Compounds},
  author = {Yoshimura, Kazuyoshi and Nakamura, Yoji},
  year = {1983},
  month = dec,
  journal = {Journal of Magnetism and Magnetic Materials},
  volume = {40},
  number = {1-2},
  pages = {55--60},
  issn = {03048853},
  doi = {10.1016/0304-8853(83)90010-0},
  urldate = {2024-09-15},
  copyright = {https://www.elsevier.com/tdm/userlicense/1.0/},
  langid = {english}
}

@article{nag2022,
  title = {Correlation Driven Near-Flat Band Stoner Excitations in a Kagome Magnet},
  author = {Nag, Abhishek and Peng, Yiran and Li, Jiemin and Agrestini, S. and Robarts, H. C. and {Garc{\'i}a-Fern{\'a}ndez}, Mirian and Walters, A. C. and Wang, Qi and Yin, Qiangwei and Lei, Hechang and Yin, Zhiping and Zhou, Ke-Jin},
  year = 2022,
  month = nov,
  journal = {Nature Communications},
  volume = {13},
  number = {1},
  pages = {7317},
  issn = {2041-1723},
  doi = {10.1038/s41467-022-34933-y},
  urldate = {2025-10-16},
  langid = {english}
}

@article{li2025,
  title = {Electron Correlation and Incipient Flat Bands in the Kagome Superconductor {CsCr}$_3${Sb}$_5$},
  author = {Li, Yidian and Liu, Yi and Du, Xian and Wu, Siqi and Zhao, Wenxuan and Zhai, Kaiyi and Hu, Yinqi and Zhang, Senyao and Chen, Houke and Liu, Jieyi and Yang, Yiheng and Peng, Cheng and Hashimoto, Makoto and Lu, Donghui and Liu, Zhongkai and Wang, Yilin and Chen, Yulin and Cao, Guanghan and Yang, Lexian},
  year = {2025},
  month = apr,
  journal = {Nature Communications},
  volume = {16},
  number = {1},
  pages = {3229},
  issn = {2041-1723},
  doi = {10.1038/s41467-025-58487-x},
  urldate = {2025-10-16},
  langid = {english}
}

@article{biswas2025,
  title = {Tuning Magnetic Ground States of R Mn 6 Sn 6 ( R = Lu, Mg) Kagome Metals by Dimensionality Reduction: Route to Ferromagnetism and Large Anomalous Hall Effect},
  shorttitle = {Tuning Magnetic Ground States of R Mn 6 Sn 6 ( R = Lu, Mg) Kagome Metals by Dimensionality Reduction},
  author = {Biswas, Rajdeep and Sharma, Jyoti and Alam, Aftab and Dasgupta, Tanusri Saha},
  year = 2025,
  month = oct,
  journal = {Physical Review Materials},
  volume = {9},
  number = {10},
  pages = {104205},
  issn = {2475-9953},
  doi = {10.1103/h9ym-4cp2},
  urldate = {2026-01-03},
  langid = {english}
}

@misc{supp,
  note = "See Supplemental Material at
    URL-will-be-inserted-by-publisher."
}

@article{kang2020b,
  title = {Topological Flat Bands in Frustrated Kagome Lattice {CoSn}},
  author = {Kang, Mingu and Fang, Shiang and Ye, Linda and Po, Hoi Chun and Denlinger, Jonathan and Jozwiak, Chris and Bostwick, Aaron and Rotenberg, Eli and Kaxiras, Efthimios and Checkelsky, Joseph G. and Comin, Riccardo},
  year = {2020},
  month = aug,
  journal = {Nature Communications},
  volume = {11},
  number = {1},
  pages = {4004},
  issn = {2041-1723},
  doi = {10.1038/s41467-020-17465-1},
  urldate = {2024-09-14},
  langid = {english}
}

@article{yin2022,
  title = {Topological Kagome Magnets and Superconductors},
  author = {Yin, Jia Xin and Lian, Biao and Hasan, M. Zahid},
  year = {2022},
  month = dec,
  journal = {Nature},
  volume = {612},
  number = {7941},
  pages = {647--657},
  publisher = {Nature Research},
  issn = {14764687},
  doi = {10.1038/s41586-022-05516-0},
  pmid = {36543954}
}

@article{sales2022,
  title={Chemical Control of Magnetism in the Kagome Metal {CoSn}$_{1-x}${In}$_x$: Magnetic Order from Nonmagnetic Substitutions},
  author={Sales, Brian C and Meier, William R and Parker, David S and Yin, Li and Yan, Jiaqiang and May, Andrew F and Calder, Stuart and Aczel, Adam A and Zhang, Qiang and Li, Haoxiang and Yilmaz, Turgut and Vescovo, Elio and Miao, Hu and Moseley, Duncan H. and Hermann, Raphael P.  and McGuire, Michael A.},
  journal={Chemistry of Materials},
  volume={34},
  number={15},
  pages={7069--7077},
  year={2022},
  publisher={ACS Publications}
}

@article{ma2022,
  title={Magnetic and charge instabilities in vanadium-based topological kagome metals},
  author={Ma, Hai-Yang and Yin, Jia-Xin and Zahid Hasan, M and Liu, Jianpeng},
  journal={Physical Review B},
  volume={106},
  number={15},
  pages={155125},
  year={2022},
  publisher={APS}
}

@article{Roychowdhury2022,
  title = {Large Room Temperature Anomalous Transverse Thermoelectric Effect in Kagome Antiferromagnet {YMn$_6$Sn$_6$}},
  author = {Roychowdhury, Subhajit and Ochs, Andrew M. and Guin, Satya N. and Samanta, Kartik and Noky, Jonathan and Shekhar, Chandra and Vergniory, Maia G. and Goldberger, Joshua E. and Felser, Claudia},
  year = {2022},
  month = oct,
  journal = {Advanced Materials},
  volume = {34},
  number = {40},
  pages = {2201350},
  issn = {0935-9648, 1521-4095},
  doi = {10.1002/adma.202201350},
  urldate = {2024-09-15},
  langid = {english}
}

@article{Siegfried2022,
  title = {Magnetization-Driven Lifshitz Transition and Charge-Spin Coupling in the {K}agome Metal {YMn}$_6${Sn}$_6$},
  author = {Siegfried, Peter E. and Bhandari, Hari and Jones, David C. and Ghimire, Madhav P. and Dally, Rebecca L. and Poudel, Lekh and Bleuel, Markus and Lynn, Jeffrey W. and Mazin, Igor I. and Ghimire, Nirmal J.},
  year = {2022},
  month = mar,
  journal = {Communications Physics},
  volume = {5},
  number = {1},
  pages = {58},
  issn = {2399-3650},
  doi = {10.1038/s42005-022-00833-2},
  urldate = {2024-09-14},
  langid = {english}
}

@article{Lichtenstein1998,
   author = {A I Lichtenstein and M I Katsnelson},
   doi = {10.1103/PhysRevB.57.6884},
   isbn = {0163-1829},
   issn = {0163-1829},
   issue = {12},
   journal = {Phys. Rev. B},
   pages = {6884},
   publisher = {American Physical Society},
   title = {Ab initio calculations of quasiparticle band structure in correlated systems: {LDA}++ approach},
   volume = {57},
   url = {http://link.aps.org/doi/10.1103/PhysRevB.57.6884 http://arxiv.org/abs/cond-mat/9707127},
   year = {1997}
}

@article{Anisimov2012,
   author = {V. I. Anisimov and A. S. Belozerov and A. I. Poteryaev and I. Leonov},
   doi = {10.1103/PhysRevB.86.035152},
   issn = {1098-0121},
   issue = {3},
   journal = {Physical Review B},
   month = {7},
   pages = {035152},
   title = {Rotationally invariant exchange interaction: The case of paramagnetic iron},
   volume = {86},
   url = {http://link.aps.org/doi/10.1103/PhysRevB.86.035152},
   year = {2012}
}

@article{Horiba2004,
  title = {Nature of the Well Screened State in Hard X-Ray {Mn} $2p$ Core-Level Photoemission Measurements of {La}$_{1-x}${Sr}$_x${MnO}$_3$ Films},
  author = {Horiba, K. and Taguchi, M. and Chainani, A. and Takata, Y. and Ikenaga, E. and Miwa, D. and Nishino, Y. and Tamasaku, K. and Awaji, M. and Takeuchi, A. and Yabashi, M. and Namatame, H. and Taniguchi, M. and Kumigashira, H. and Oshima, M. and Lippmaa, M. and Kawasaki, M. and Koinuma, H. and Kobayashi, K. and Ishikawa, T. and Shin, S.},
  journal = {Phys. Rev. Lett.},
  volume = {93},
  issue = {23},
  pages = {236401},
  numpages = {4},
  year = {2004},
  month = {Nov},
  publisher = {American Physical Society},
  doi = {10.1103/PhysRevLett.93.236401},
  url = {https://link.aps.org/doi/10.1103/PhysRevLett.93.236401}
}

@article{Uozumi1997,
title = {Theoretical and experimental studies on the electronic structure of {M}$_2${O}$_3$ ({M} = {Ti}, {V}, {Cr}, {Mn}, {Fe}) compounds by systematic analysis of high-energy spectroscopy},
journal = {Journal of Electron Spectroscopy and Related Phenomena},
volume = {83},
number = {1},
pages = {9-20},
year = {1997},
issn = {0368-2048},
doi = {https://doi.org/10.1016/S0368-2048(96)03063-0},
url = {https://www.sciencedirect.com/science/article/pii/S0368204896030630},
author = {T. Uozumi and K. Okada and A. Kotani and R. Zimmermann and P. Steiner and S. H\"{u}fner and Y. Tezuka and S. Shin}
}

@article{Hariki2018,
  title = {{Continuum Charge Excitations in High-Valence Transition-Metal Oxides Revealed by Resonant Inelastic X-Ray Scattering}},
  author = {Hariki, Atsushi and Winder, Mathias and Kune\ifmmode \check{s}\else \v{s}\fi{}, Jan},
  journal = {Phys. Rev. Lett.},
  volume = {121},
  issue = {12},
  pages = {126403},
  numpages = {6},
  year = {2018},
  month = {Sep},
  publisher = {American Physical Society},
  doi = {10.1103/PhysRevLett.121.126403},
  url = {https://link.aps.org/doi/10.1103/PhysRevLett.121.126403}
}

@article{Zhou2011,
  title = {{Localized and delocalized Ti 3$d$ carriers in LaAlO${}_{3}$/SrTiO${}_{3}$ superlattices revealed by resonant inelastic x-ray scattering}},
  author = {Zhou, Ke-Jin and Radovic, Milan and Schlappa, Justine and Strocov, Vladimir and Frison, Ruggero and Mesot, Joel and Patthey, Luc and Schmitt, Thorsten},
  journal = {Phys. Rev. B},
  volume = {83},
  issue = {20},
  pages = {201402},
  numpages = {4},
  year = {2011},
  month = {May},
  publisher = {American Physical Society},
  doi = {10.1103/PhysRevB.83.201402},
  url = {https://link.aps.org/doi/10.1103/PhysRevB.83.201402}
}

@article{Bisogni2016,
  title={Ground-state oxygen holes and the metal--insulator transition in the negative charge-transfer rare-earth nickelates},
  author={Bisogni, Valentina and Catalano, Sara and Green, Robert J and Gibert, Marta and Scherwitzl, Raoul and Huang, Yaobo and Strocov, Vladimir N and Zubko, Pavlo and Balandeh, Shadi and Triscone, Jean-Marc and Sawatzky, George and Schmitt, Thorsten},
  journal={Nature Communications},
  volume={7},
  number={1},
  pages={13017},
  year={2016},
  publisher={Nature Publishing Group UK London}
}

@article{Yamauchi2025,
author = {Yamaguchi, Daiki and Kitaori, Aki and Nagaosa, Naoto and Tokura, Yoshinori},
title = {Current Control of Spin Helicity and Nonreciprocal Charge Transport in a Multiferroic Conductor},
journal = {Advanced Materials},
volume = {37},
number = {20},
pages = {2420614},
doi = {https://doi.org/10.1002/adma.202420614},
year = {2025}
}

@article{Ghiringhelli2007,
  title = {{Sensitivity to hole doping of Cu ${L}_{3}$ resonant spectroscopies: Inelastic x-ray scattering and photoemission of ${\mathrm{La}}_{2\ensuremath{-}x}{\mathrm{Sr}}_{x}\mathrm{Cu}{\mathrm{O}}_{4}$}},
  author = {Ghiringhelli, G. and Brookes, N. B. and Dallera, C. and Tagliaferri, A. and Braicovich, L.},
  journal = {Phys. Rev. B},
  volume = {76},
  issue = {8},
  pages = {085116},
  numpages = {7},
  year = {2007},
  month = {Aug},
  publisher = {American Physical Society},
  doi = {10.1103/PhysRevB.76.085116},
  url = {https://link.aps.org/doi/10.1103/PhysRevB.76.085116}
}

@article{Gilmore2021,
  title = {{Description of Resonant Inelastic X-Ray Scattering in Correlated Metals}},
  author = {Gilmore, Keith and Pelliciari, Jonathan and Huang, Yaobo and Kas, Joshua J. and Dantz, Marcus and Strocov, Vladimir N. and Kasahara, Shigeru and Matsuda, Yuji and Das, Tanmoy and Shibauchi, Takasada and Schmitt, Thorsten},
  journal = {Phys. Rev. X},
  volume = {11},
  issue = {3},
  pages = {031013},
  numpages = {24},
  year = {2021},
  month = {Jul},
  publisher = {American Physical Society},
  doi = {10.1103/PhysRevX.11.031013},
  url = {https://link.aps.org/doi/10.1103/PhysRevX.11.031013}
}

@article{Ghiringhelli2006,
  title = {{Resonant inelastic x-ray scattering of $\mathrm{MnO}$: ${\mathrm{L}}_{2,3}$ edge measurements and assessment of their interpretation}},
  author = {Ghiringhelli, G. and Matsubara, M. and Dallera, C. and Fracassi, F. and Tagliaferri, A. and Brookes, N. B. and Kotani, A. and Braicovich, L.},
  journal = {Phys. Rev. B},
  volume = {73},
  issue = {3},
  pages = {035111},
  numpages = {4},
  year = {2006},
  month = {Jan},
  publisher = {American Physical Society},
  doi = {10.1103/PhysRevB.73.035111},
  url = {https://link.aps.org/doi/10.1103/PhysRevB.73.035111}
}

@article{Perfetti2003,
  title = {{Spectroscopic Signatures of a Bandwidth-Controlled Mott Transition at the Surface of $1T\mathrm{\text{\ensuremath{-}}}{\mathrm{T}\mathrm{a}\mathrm{S}\mathrm{e}}_{2}$}},
  author = {Perfetti, L. and Georges, A. and Florens, S. and Biermann, S. and Mitrovic, S. and Berger, H. and Tomm, Y. and H\"ochst, H. and Grioni, M.},
  journal = {Phys. Rev. Lett.},
  volume = {90},
  issue = {16},
  pages = {166401},
  numpages = {4},
  year = {2003},
  month = {Apr},
  publisher = {American Physical Society},
  doi = {10.1103/PhysRevLett.90.166401},
  url = {https://link.aps.org/doi/10.1103/PhysRevLett.90.166401}
}

@article{Takizawa2009,
  title = {Coherent and incoherent $d$ band dispersions in {SrVO}$_{3}$},
  author = {Takizawa, M. and Minohara, M. and Kumigashira, H. and Toyota, D. and Oshima, M. and Wadati, H. and Yoshida, T. and Fujimori, A. and Lippmaa, M. and Kawasaki, M. and Koinuma, H. and Sordi, G. and Rozenberg, M.},
  journal = {Phys. Rev. B},
  volume = {80},
  issue = {23},
  pages = {235104},
  numpages = {4},
  year = {2009},
  month = {Dec},
  publisher = {American Physical Society},
  doi = {10.1103/PhysRevB.80.235104},
  url = {https://link.aps.org/doi/10.1103/PhysRevB.80.235104}
}

@article{DJH2003,
  title = {{Anomalous spin polarization and dualistic electronic nature of ${\mathrm{CrO}}_{2}$}},
  author = {Huang, D. J. and Tjeng, L. H. and Chen, J. and Chang, C. F. and Wu, W. P. and Chung, S. C. and Tanaka, A. and Guo, G. Y. and Lin, H.-J. and Shyu, S. G. and Wu, C. C. and Chen, C. T.},
  journal = {Phys. Rev. B},
  volume = {67},
  issue = {21},
  pages = {214419},
  numpages = {5},
  year = {2003},
  month = {Jun},
  publisher = {American Physical Society},
  doi = {10.1103/PhysRevB.67.214419},
  url = {https://link.aps.org/doi/10.1103/PhysRevB.67.214419}
}

@article{Fujimori1992,
  title = {Evolution of the spectral function in Mott-Hubbard systems with ${\mathit{d}}^{1}$ configuration},
  author = {Fujimori, A. and Hase, I. and Namatame, H. and Fujishima, Y. and Tokura, Y. and Eisaki, H. and Uchida, S. and Takegahara, K. and de Groot, F. M. F.},
  journal = {Phys. Rev. Lett.},
  volume = {69},
  issue = {12},
  pages = {1796--1799},
  numpages = {0},
  year = {1992},
  month = {Sep},
  publisher = {American Physical Society},
  doi = {10.1103/PhysRevLett.69.1796},
  url = {https://link.aps.org/doi/10.1103/PhysRevLett.69.1796}
}
\end{document}